\documentclass[pra,twocolumn,superscriptaddress,showpacs,amsmath,amssymb]{revtex4}

\usepackage{bm}

\usepackage{graphicx}
\usepackage{latexsym}
\usepackage{amsmath}
\usepackage{amssymb}
\usepackage{amsfonts}
\usepackage{color}
\usepackage{verbatim}
\usepackage{epstopdf}

\begin{document}

\title{Lifshitz transition in the double-core vortex in $^3$He-B}


\author{M.\,A.\,Silaev}%
\affiliation{Department of Theoretical Physics, KTH-Royal
Institute of Technology, SE-10691 Stockholm,  Sweden}


\author{E.\,V.\,Thuneberg}%
 \affiliation{Department of Physics, University of Oulu, FI-90014, Finland}

\author{M. Fogelstr{\"o}m }%
 \affiliation{Department of Microtechnology and Nanoscience, Chalmers, SE-41296 G\"oteborg, Sweden}


\date{\today}

\begin{abstract}

We study the spectrum of fermion states localized within the
vortex core of a weak-coupling p-wave superfluid. The low energy
spectrum consists of two anomalous branches that generate large
density of states at the locations of the half cores of the
vortex. Fermi liquid interactions significantly stretch the vortex
structure, which leads to Lifshitz transition in the effective Fermi surface of the vortex core fermions. We apply the results to rotational
dynamics of vortices in superfluid $^3$He-B and find explanation
for the observed slow mode.
\end{abstract}


 \maketitle

\begin{figure}[tb]
 \centerline{$
 \begin{array}{c}
  \includegraphics[width=1.0\linewidth]{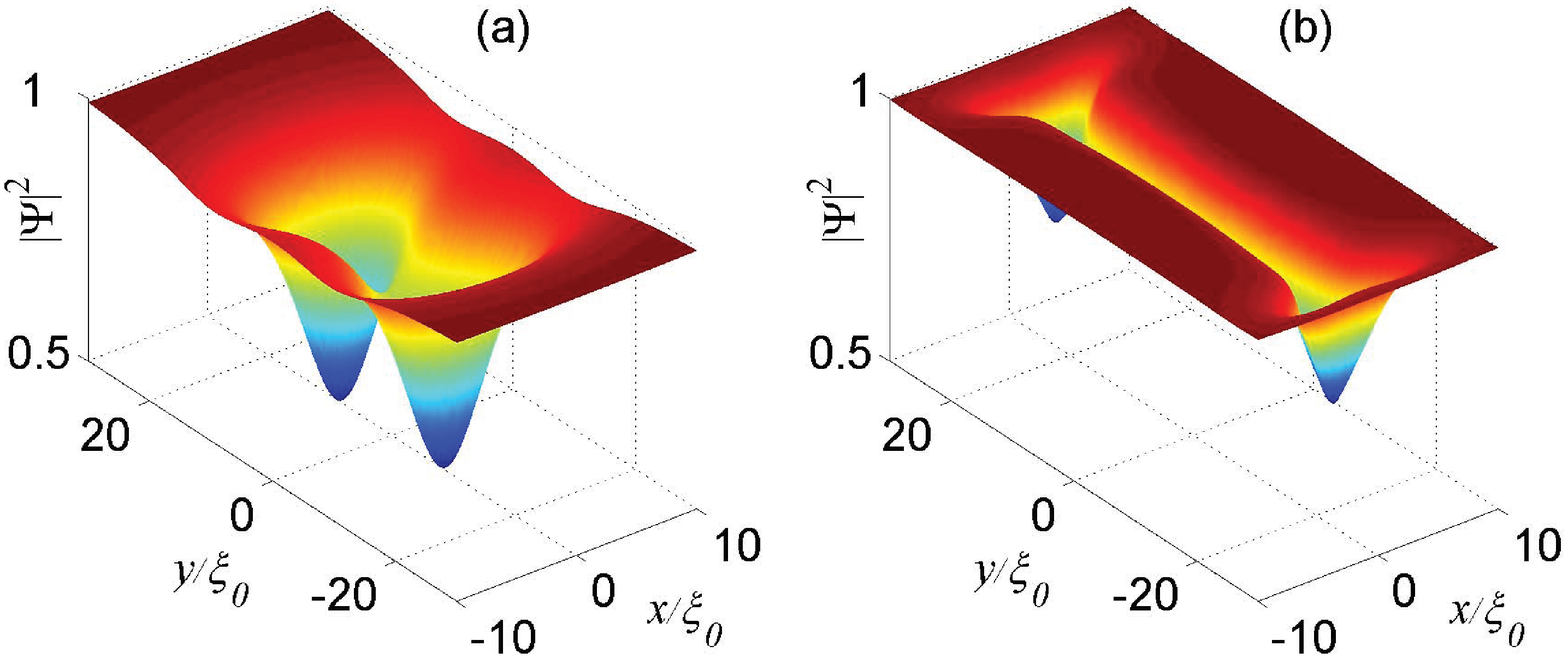}\\
  \includegraphics[width=1.0\linewidth]{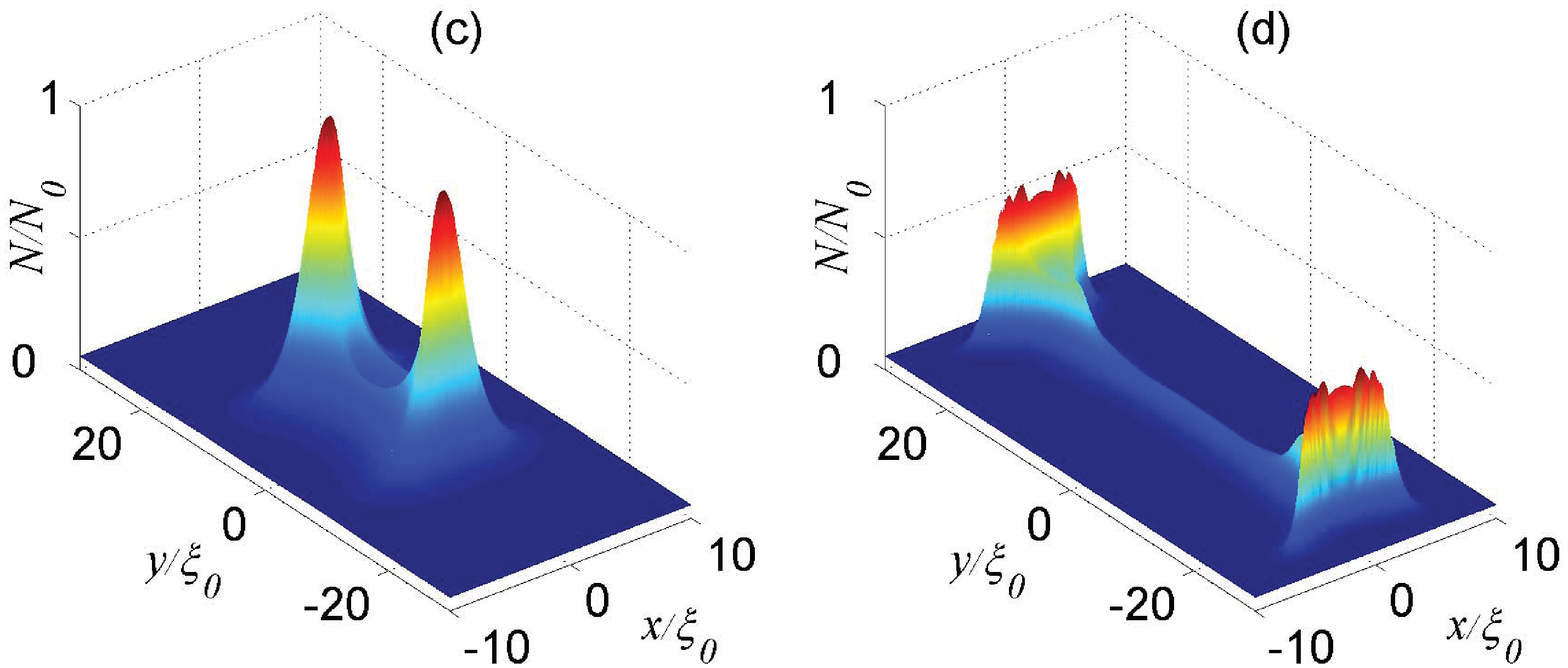}
 \end{array}$}
 \caption{\label{Fig:LDOS3D} (Color online)
 (a,b) The double-core vortex structure made visible by  the pair density  $|\Psi|^2 = \sum_{\mu,i}|A_{\mu,i}|^2$
plotted in the $x-y$ plane perpendicular to the vortex axis at
temperature $T/T_c=0.9$, and (b) $T/T_c=0.1$.
 (c,d) The normalized local density of states profiles demonstrating the
 quasiparticle wave function at the Fermi level $\varepsilon=0$, $\hat p_z=0$ at $T/T_c=0.9$ and (d)
 $T/T_c=0.1$. All plots correspond to pressure $P= 24$ bar.}
 \end{figure}

The double-core vortex is an amazing structure because it is the unique answer
to a simple question:
 what is the vortex structure of a weak-coupling p-wave-pairing superfluid.
The ground state in this case is  the Balian-Werthamer (BW) state
\cite{Balian63}. The lowest energy vortex has the double-core
structure, where the core is split into two ``half cores'' as
depicted in Fig.\ \ref{Fig:LDOS3D}(a,b)
\cite{Thuneberg-PRL,VolovikSalomaa-PRL,Thuneberg-PRB,Volovik90,Fogelstrom-1,Fogelstrom-2,Kita02}.
This is not only of theoretical interest since superfluid $^3$He
is close to being weak coupling, and its B phase was identified as
the BW state. Two vortex types have been found experimentally  in
$^3$He-B \cite{experiment1,experiment2,experiment3,BevanJLTP1997}.
The vortex being stable in the major, low-pressure part of the
phase diagram  was identified as the double-core vortex. The
vortex stable at higher pressures was identified as the axially
symmetric A-phase-core vortex. Available experimental evidence is
consistent with the theoretical identification of the vortex
structures. In particular, the broken axial symmetry of the
double-core vortex was used to explain the peculiar dynamical
properties that have been observed for the low pressure vortex
using homogeneous precessing domain (HPD)  mode of  NMR
\cite{Dmitriev90,Kondo91}. A similar double-core
vortex structure has been suggested to appear in spin-triplet
heavy fermion superconductor UPt$_3$ \cite{TsutsumiUPt3}.

One of the most interesting properties of quantized vortices in
superconductors and Fermi superfluids is the presence of fermionic
quasiparticles localized within vortex core at energies smaller
than the bulk energy gap \cite{CdGM, BardeenJacobs}. Generally
fermionic bound states determine both thermodynamic and dynamic
properties of vortices at low temperatures
\cite{LarkinOvchinnikovImpurity,RainerSauls,StoneMgnusForce,KopninReview,MachidaSpecificHeat}.
In the rotational dynamics of the double-core vortex they are
predicted to give rise to resonance absorption at the frequency
comparable with the spacing of the localized energy eigenstates
\cite{RotatingVortex}. Recently much attention has been focused on
the topologically protected zero energy vortex-core and surface
states in superfluid $^3$He
\cite{Silaev10,Silaev14,Mizushima15,Tsutsumi15}. Particularly
motivating is a predicted existence of self-conjugated Majorana
states localized on half-quantum vortices in p-wave superfluids
\cite{IvanovPRL}.

In this letter we calculate the low-energy fermionic excitation
spectrum of the double-core vortex. We find that the low-energy
excitations mostly are localized in the two half cores.
 This is visualized in
Fig.\ \ref{Fig:LDOS3D}(c,d) which show the fermionic local density
of states (LDOS) profiles around the vortex core. We can interpret
the two half cores as potential wells for quasiparticles. The
motion of the excitations between the wells depends on the
potential barrier between them. We find that this barrier changes
essentially as the distance of the wells changes as a function of
pressure and temperature
 (see Fig.\ \ref{Fig:LDOS3D} to compare vortex structures at $T/T_c = 0.9$ and $T/T_c
 =0.1$). This implies a transition from
excitations that circle both half cores to separate excitations
 that circle only a single half core. 
This transition can be seen as a
Lifshitz-type transition in the topology of the fermionic states
bound to the vortex core.  We
discuss how this could be observable in the time scales of
rotational dynamics. Comparing our calculations with an earlier
experiment reveals a serious disagreement in the model used to
interpret the experimental data \cite{Kondo91}. We construct a
different model, which also provides an explanation for the long
time scale observed in rotational dynamics
\cite{Dmitriev90,Sonin93,Krusius93}.

 The triplet pairing of fermions in orbital $p$-wave states is described by the matrix
  \begin{equation}\label{OPgen}
 \check\Delta({\bm r}, \hat{\bm p})=\sum_{\alpha, i}A_{\alpha i} ({\bm r})i \check{\bm
 \sigma}_\alpha\check\sigma_y\hat p_i \ ,
 \end{equation}
 where $ \check\sigma_{x, y, z}$ are Pauli matrices,  $\bm p$
 is the momentum close to the Fermi surface $p\approx p_F=\hbar k_F$, and $\hat {\bm p}=\bm p/p$.
 The gap function $ \check\Delta$ \eqref{OPgen} is determined by the $3\times 3$ order
 parameter matrix with complex components $A_{\alpha i}$. Here $\alpha=x, y, z$ and $i=x, y, z$
 are spin and orbital indices, respectively.

 In the weak coupling theory of $p$-wave superfluid, the stable state has
 the Balian-Werthamer (BW) form \cite{Balian63}.
In BW state, the order parameter far from the vortex axis is
$A_{\alpha i}=\Delta_0 \exp(i\varphi) R_{\alpha i}$. Here
$\varphi$ is the azimuth with respect to the vortex axis,
$R_{\alpha i}$ is a constant rotation matrix and $\Delta_0$ the
order parameter amplitude. Near the vortex axis a more
sophisticated structure appears
\cite{Thuneberg-PRL,VolovikSalomaa-PRL,Thuneberg-PRB,Volovik90,Fogelstrom-1,Fogelstrom-2,Kita02}.
 It is energetically favorable to change the sign of the order
 parameter across the vortex axis by spin rotation of the BW-state matrix $A_{\alpha i}$ by $\pi$ \cite{Thuneberg-PRB}.
 This effectively results in splitting of a singly-quantized
 vortex to a pair of half-quantum vortices that are  bound together by a
 planar-phase domain wall. For illustration see 
  Figs.\ \ref{Fig:LDOS3D}(a) and \ref{Fig:LDOS3D}(b), which
  show the pair density $|\Psi|^2 = \sum_{\alpha, i}|A_{\alpha i}|^2 $
  in the $x-y$ plane. The pair density has two distinct minima, whence the name double-core vortex.

To determine the vortex structure we calculate self-consistently
the order parameter and the Fermi-liquid self
energy \cite{Rainer}. The numerics is performed as described in
Ref.\ \onlinecite{Fogelstrom-1}, i.e.\ using the explosion trick
to solve the Eilenberger transport equation. We extend the
previous work \cite{Fogelstrom-1,Fogelstrom-2} to higher accuracy,
lower temperatures and different values of the Fermi-liquid
parameter $F_1^s$ corresponding to different pressures. The
parameter $F_1^s$ determines the feedback of superfluid mass current
on the order parameter and can significantly change
both the vortex structure and spectrum of bound fermions.

\begin{figure}[tb!]
 \centerline{\includegraphics[width=0.7\linewidth]{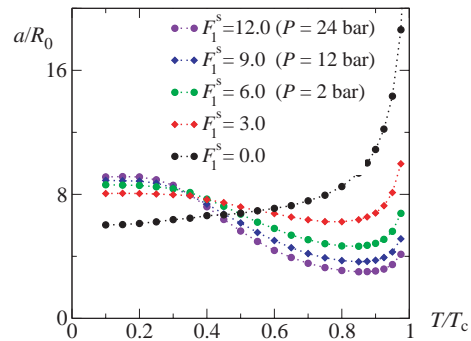}}
 \caption{\label{Fig:OrderParameterFermi} (Color online)
Distance $a$ between the half cores  in the double-core vortex at
different values of the Fermi
 liquid parameter $F_{1}^s$. The locations of the half cores are determined from zeros of supercurrent density.}
 \end{figure}

 The distance $a$ between the half cores  is shown
Fig.\ \ref{Fig:OrderParameterFermi}. Its scale is $R_0=\left(1+F_1^s/3\right)\xi_0$,
where $\xi_0=\hbar v_F/2\pi T_c$ is the coherence length
and $v_F$ the Fermi velocity.  As $F_1^{\rm s}$ in liquid $^3$He ranges from 5.4
to 14.6 depending on pressure $P$ \cite{Greywall86}, the two length scales can differ essentially.
Thus at large values of  $F_1^{\rm s}$, corresponding to high pressures, the vortex
size at a low temperature is much larger than the coherence length.
For example, $a = 46\xi_0$ in the case of
Fig.\ \ref{Fig:LDOS3D}(b).
Fig.\ \ref{Fig:OrderParameterFermi} also shows strong temperature and pressure dependence.  The distance of the half cores
 grows almost 3-fold when the temperature decreases from $0.9
 T_c$ to $0.1 T_c$ at $24$ bar. Similarly as a function of
 pressure the distance $a$ measured in units of $\xi_0$,
 grows almost 2-fold when the pressure increases from $0$  to $24$ bar at a low
 temperature.

We calculate the excitation spectrum of the double-core vortex
using the self-consistent order parameter field. This is done by
solving the eigenvalue problem for the system of Andreev equations,
which are ordinary differential equations describing the
propagation of quasiparticle wave function along classical
trajectories.

\begin{figure}[tb]
 \centerline{$
 \begin{array}{c}
 \includegraphics[width=0.5\linewidth]{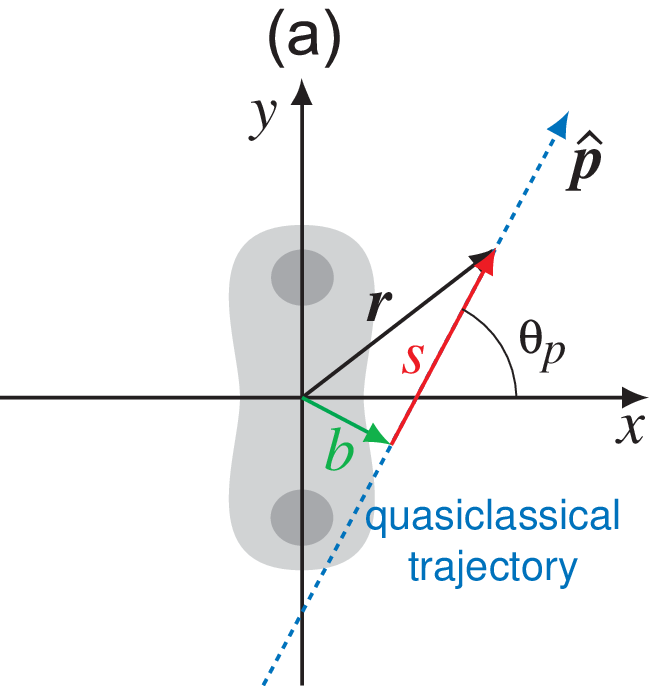}
 \includegraphics[width=0.5\linewidth]{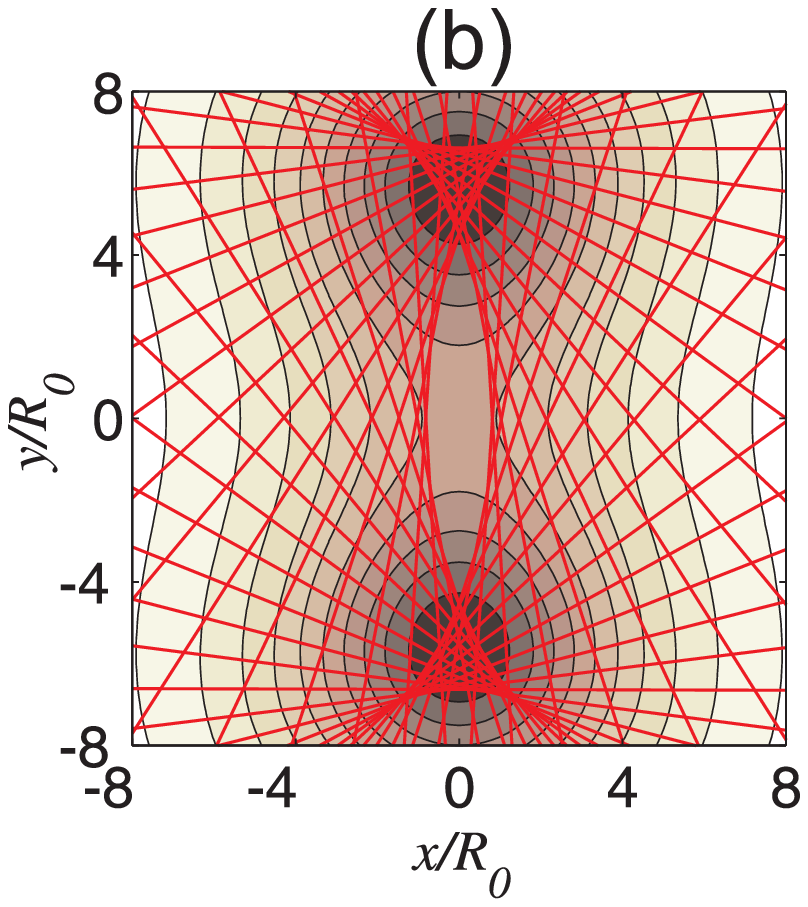} \\
 \includegraphics[width=1.0\linewidth]{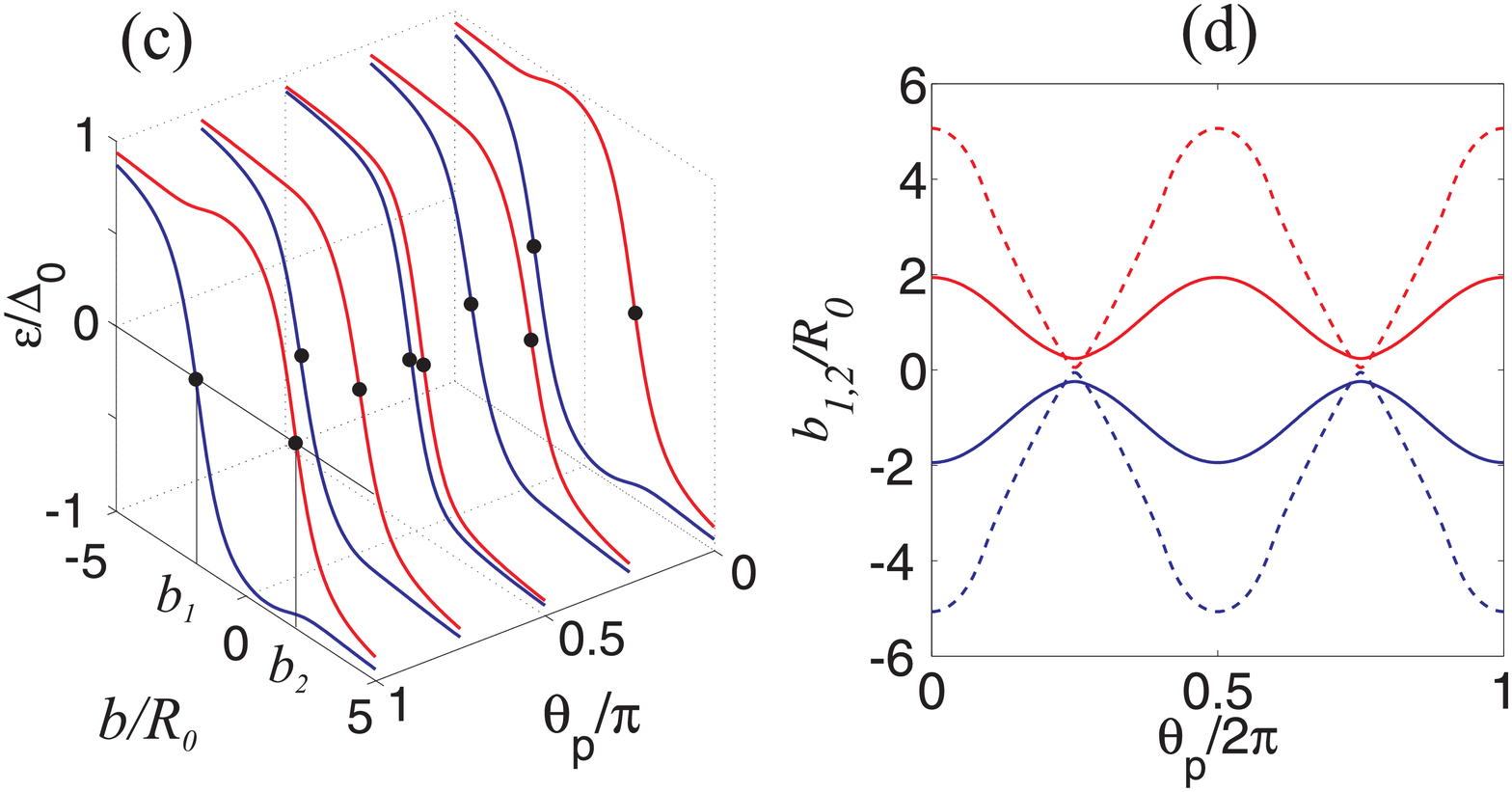}
 \end{array}$}
 \caption{\label{Fig:Spectrum} (Color online)
   (a) Schematic plot of a quasiclassical
 trajectory in the $x-y$ plane in the direction of $(\cos\theta_p, \sin\theta_p)$
 passing the vortex axis at  distance $b$ (impact parameter). A point ${\bm r}=(x,y)$ on the
 trajectory is determined by the coordinate $s$.
 (b) A bunch of quasiparticle trajectories (straight lines) at the Fermi level $\varepsilon=0$ and
 $\hat p_z=0$ superimposed on the pair-density contour plot at
 $T/T_c=0.5$ and $F_1^s=0$. (c) Anomalous branches of the quasiparticle
 spectrum $\varepsilon=\varepsilon_{1,2} (b,\theta_p)$ at $\hat p_z=0$, $F_1^s=12$ and $T=0.9 T_c$.
 (d) Two sheets of effective Fermi surface $b=b_{1, 2}(\theta_p)$ at $\hat p_z=0$,  $F_1^s=12$,  $T=0.9 T_c$ (solid lines)
 and $T=0.1 T_c$ (dashed lines).  }
 \end{figure}

The momentum $\bm p$ of a low energy excitation is close to the
Fermi surface, $p\approx p_F$. The classical trajectories are
straight lines parallel to $\bm p$. In studying a vortex we fix
the $z$ axis as the vortex axis, and we  parameterize the momentum
direction $ \bm \hat{\bm p}=(\hat p_\perp \cos\theta_p, \hat
p_\perp\sin\theta_p, \hat p_z)$. The direction on the trajectory
is fixed by  giving $\hat p_z$ and $\theta_p$. The location of the
trajectory is given by the impact parameter $b$, the coordinate
measuring the distance from the vortex axis. The parameterization
is visualized in Fig.~\ref{Fig:Spectrum}(a). The impact parameter
is related to the projection of the angular momentum $\mu$ on the
vortex axis through the usual classical mechanics formula $\mu=p
_\perp b$. The quasiclassical   energy spectrum is  given by
$\varepsilon=\varepsilon_{i} (\hat p_z, \theta_p, b)$, where the
parameters $\hat p_z,\theta_p,b$ specify the classical trajectory
and integer $i$ counts the eigenvalues of Andreev equations on a
given trajectory \cite{AndreevEqs}.
 Figure \ref{Fig:Spectrum}(b) shows a bunch of trajectories at the Fermi level and $\hat p_z=0$.
 The concentration of the trajectories at the two half cores results in the large LDOS at the half cores.
The concave triangular shape of the caustic of the trajectories at
the half cores is clearly visible in the LDOS shown in Fig.\
\ref{Fig:LDOS3D}(d). Also the classically non-allowed region
around the vortex axis in Fig.\ \ref{Fig:Spectrum}(b) can be
recognized in Fig.\ \ref{Fig:LDOS3D}(c) as a valley in the
LDOS profile in the region between the half cores.

 Due to the lifted spin degeneracy, singly-quantized vortices in $^3$He-B have two
anomalous branches of quasiparticle spectrum \cite{SilaevJETPlett}.
At low energy compared to the bulk energy gap, $|\varepsilon|\ll
\Delta_0$, they can be represented as
\begin{equation}\label{LowEnergySpectrum-gen}
\varepsilon_i (\hat p_z, \theta_p, b)=-\omega_{i} p_F(b - b_{i}),
\end{equation}
where $i=1,2$. Here $b_i
(\hat p_z,\theta_p )$ is the impact parameter that corresponds to vanishing excitation
energy, and  $\omega_{i}( \hat p_z,\theta_p)$ indicates the slope of the energy at $b=b_i$.
The new feature in a non-axisymmetric vortex is that these parameters depend on the trajectory direction
$\theta_p$ in the $x-y$ plane.

 Fig.~\ref{Fig:Spectrum}(c) shows the calculated
 quasiparticle energies as a function of impact parameter $b$ and different directions of the trajectory
 $\theta_p$.
 The curves cross the Fermi level at a finite $b$ in
 accordance with Eq.~\eqref{LowEnergySpectrum-gen}.
These locations in  the energy spectrum $b_{1,2}
 (\theta_p)$ are shown by the black dots in Fig.~\ref{Fig:Spectrum}(c).
  The states at the Fermi level in the spectrum \eqref{LowEnergySpectrum-gen} form a 2D {\it effective Fermi surface}
  $b=b_{1,2}(\hat p_z,\theta_p)$ in the 3D space formed by
 the quasiclassical quantum numbers ($\hat p_z, \theta_p, b$) in the vortex core. Because of two nondegenerate branches
 \eqref{LowEnergySpectrum-gen}, there are two sheets in the Fermi surface.
 One more representation of this is given in Fig.\ \ref{Fig:Spectrum}(d).
 It shows  $b_{1,2}$ as a function of $ \theta_p$. The curves depend also on $\hat p_z$ but that dependence is less important
 in the following because $\hat p_z$ is conserved. For comparison, the trajectories passing precisely through the half cores
 at $y=\pm a/2$ would correspond to curves $b(\theta_p)=\mp\frac12a\cos\theta_p$.

The topology of the effective Fermi surface is determined by the
behavior of
 zero energy lines $b_{1,2}(\theta_p)$
 at $\theta_p=\pi(n+1/2)$ with $n=1, 2$. At these angles the quasiparticle trajectories pass through both half cores.
 In general there is overlap of the quasiparticle wave functions localized at different half cores. This makes that there
 is no sign change of  $b_{1,2}(\theta_b)$. That is, there is anticrossing of the two branches  and a finite splitting
 $2\delta b=|b_1-b_2|>0$ at $\theta_p=\pi/2$, as shown by the solid lines in Fig.\ \ref{Fig:Spectrum}(d). Physically this
  means that an excitation created at one half core will jump periodically between the half cores.

The growing core separation (compared to $\xi_0$) at  low
temperatures and large pressures reduces the overlap of the
 quasiparticle wave functions located at different half cores. As a result the splitting $2\delta b$ becomes extremely small
as shown by dashed lines in Fig.\ \ref{Fig:Spectrum}(d) for $F_1^s=12$ and $T=0.1 T_c$.
 In this case {\it Landau-Zener} (LZ) tunneling between the quasiclassical branches (\ref{LowEnergySpectrum-gen}) becomes
 important. The probability $W$ of these transitions can be found from
 the conventional approach \cite{LandauLifshitzQM,MelnikovRyzhovSilaev-Meso} by taking $\theta_p$ and  the angular momentum $\mu=p_\perp b$
 as the conjugate variables. Near the anticrossing point at $\theta_p=\pi/2$ we can approximate
$b_{1,2}(\theta_p)  \approx \pm \sqrt{\delta b^2 + (a\theta/2)^2}$, where $\theta=\theta_p-\pi/2$.
The transition probability is given by
 $W = \exp [ -2k_\perp {\rm Im} \int_0^{i\theta^*} (b_1-b_2) d\theta]$  where $i\theta^* = 2i\delta b/a $
 is the intersection point of quasiclassical branches in the complex plane.
 A simple calculation yields $W = \exp[-2\pi \hat p_\perp (\delta b/ \Delta b)^2 ]$, where
 $\Delta b= \sqrt{a/k_F}$ has a physical meaning of the quantum mechanical uncertainty of the impact parameter.

Once the transition probability becomes large, $W\approx 1$, the
LZ tunneling changes the topology of the effective Fermi surface
so that quasiparticles remain localized in one or the other of the
half cores.
 In Fig.\ \ref{Fig:Spectrum}(d) this means transition to intersecting zero energy curves
 $\tilde{b}_{1(2)} (\theta_p)= b_{1(2)}(\theta_p) $ for
 $-\pi/2<\theta_p<\pi/2$ and  $\tilde{b}_{1(2)} (\theta_p)= b_{2(1)}(\theta_p)$ for $\pi/2<\theta_p<3\pi/2$.
 The calculated LZ probability $W (T,P,\hat p_z)$ is shown in supplemental material \cite{SM} to demonstrate
that the condition $W\approx 1$  is realized in the double-core vortex at large pressures  and low temperatures.
The crossover from split $b_{1,2}(\theta_p)$ to  intersecting isoenergetic lines
 $\tilde{b}_{1,2}(\theta_p)$  is  an analog of the  {\it Lifshitz transition}
\cite{Lifshitz} changing the topology of the  Fermi surface.

The transition leads to a formation of two spatially separated
low-energy fermionic states localized at the half-cores. Whether there are Majorana states precisely at the Fermi level \cite{IvanovPRL} or not \cite{Tsutsumi15} is beyond our quasiclassical approach.
The transition affects the rotational dynamics of the double-core vortex. The
bound fermions in the core respond to oscillation of the core
orientation. A friction torque acting on a rotating vortex core
can be expressed by a friction coefficient $f= f_1
p_F(k_F\xi_0)^2$, where $f_1\sim 1$ is dimensionless and the
factor $p_F(k_F\xi_0)^2$ is determined by the density
 of quasiparticles in the vortex core. The expression for the friction torque \cite{RotatingVortex} yields
resonance  peaks in $f$ located at angular frequencies $\omega\approx n E_m/\hbar$ where $n$ is integer. Here
$E_m$ is the spacing  of quantized energy levels obtained from the quasiclassical spectrum
(\ref{LowEnergySpectrum-gen}) using the Bohr-Sommerfeld quantization rule for the angular  momentum \cite{RotatingVortex},
  $ E_{m}= \hbar\langle \omega^{-1}_{1}(\hat p_z=0)\rangle^{-1}$,  where $\langle...\rangle$ denotes the average over $\theta_p$.
The scale of the minigap is determined by $\hbar/\tau_n = (2\pi T_c)^2/v_Fp_F$, which is on the order of the quasiparticle
relaxation rate in the normal state.
The calculated values of the minigap are listed in Table
 \ref{table1}.
 \begin{table}
 \begin{center}
 \begin{tabular}{|p{2cm}||p{2cm}|p{2cm}|p{2cm}|p{2cm}}
 \hline
 & $P=2\; {\rm bar}$ &$P=12\; {\rm bar}$ &$P= 24\; {\rm bar}$
  \\ \hline\hline
 $T=0.05 T_c$  &$27\; {\rm kHz}$ &$71\; {\rm kHz}$ &$98\; {\rm kHz}$
  \\ \hline
 $T=0.5 T_c$ &$22\; {\rm kHz}$ &$65\; {\rm kHz}$ &$106\; {\rm kHz}$
  \\ \hline
  \end{tabular}
 \caption[]{Values of the minigap $ E_{m}/h$ at different pressures and temperatures.}
 \label{table1}
 \end{center}
 \end{table}

\begin{figure}[!tb]
 \centerline{\includegraphics[width=1.0\linewidth]{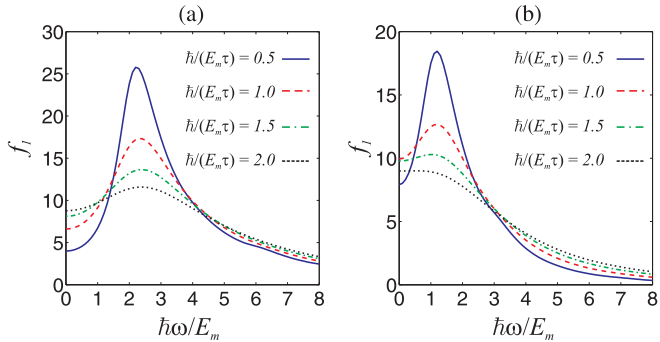}}
 \caption{\label{Fig:Friction} (Color online)
  Demonstration of the effect of the Lifshitz transition on the rotational friction coefficient $f_1$
  plotted as a function of frequency $\omega$ for (a) $W=0$,
  (b) $W=1$. The vortex structure and the excitation spectrum are calculated at $T=0.5T_c$ and
  (a) $P=2\; {\rm bar} $, (b) $P=24\; {\rm bar}$.
  The minigap values are given in Table \ref{table1}.  According to mutual friction measurements \cite{BevanJLTP1997}
  $\hbar/E_m\tau=0.7$ at  $P=24\; {\rm bar}$  but a wider range is given to illustrate
   the influence of relaxation on the shape of absorption peak. }
 \end{figure}

The amplitudes of the resonances are determined by the Fourier
amplitudes of the zero-energy curves shown in Fig.\
\ref{Fig:Spectrum}, $A_n\sim  |\int d\theta_p
e^{in\theta_p}b_1(\theta_p)|$. From the plots in Fig.\
\ref{Fig:Spectrum}(d) one can see that at  pressures below the
Lifshitz transition the largest components are those with double
frequency $\hbar\omega=2E_m$. At pressures above the transition
the amplitudes are determined by the harmonics of the intersecting
curves $\tilde{b}_i(\theta_p)$, which have strongest matrix
element at $\hbar\omega= E_m$. The difference in the  friction
coefficient $f_1$ in the two cases is demonstrated in Fig.\
\ref{Fig:Friction}. In the following we show that at least the low
frequency limit of the curves in Fig.\ \ref{Fig:Friction} is
experimentally accessible.

Kondo et al \cite{Kondo91,Dmitriev90} have studied a sample of rotating $^3$He-B using
the homogeneously precessing domain. In this mode the magnetization $\bm M$ is tipped by a large angle ($>104^\circ$)
from the field direction $\bm B$. It was found that the contribution of vortices to the relaxation changed on a few
minute time scale \cite{Dmitriev90}. This was interpreted that the double-core vortex gets twisted as its end points
(at $z=\pm L/2$) are pinned but in the bulk the rotating magnetization exerts a torque on the core.
A quantitative model was constructed for the vortex core rotation angle $\phi(t,z)$ as a function of time $t$ and  $z$.
The parameters of the model were determined by fitting to the experiment \cite{Kondo91}.
These include the friction parameter $f$, the dipole torque $T_D$, which drives the vortex in the presence of rotating magnetization,
and the rigidity $K$, which gives the energy caused by twisting the core, $F_{\rm twist}=\frac12 K (\partial_z\phi)^2$.

By precise calculation of the vortex structure we can now
calculate the vortex parameters. We find a value of $f$ that is
three orders of magnitude larger than fitted by Kondo et al
\cite{Kondo91}. Thus a serious revision of the model has to be
made.  The large value of $f$ means that only a negligible
fraction of energy dissipation comes from vortex core rotation.
Thus essentially all dissipation has to arise from
normal-superfluid disequilibrium \cite{Leggett77T}, spin
diffusion, and radiation of spin waves. Without going into
details, these can be incorporated by allowing an elastic vortex
structure, where the rotation angle $\alpha(t,z)$ at a distance
from  the vortex axis (where the dipole torque acts) can be
different from the vortex core angle $\phi(t,z)$. These are bound
by elasticity energy $\frac12T_A(\alpha-\phi)^2$, and both angles
have their own friction coefficients: $f\dot \phi=-\delta
F/\delta\phi$, $g\dot \alpha=-\delta F/\delta\alpha$. This model
results in the diffusion equation \cite{SM}
\begin{eqnarray}
\dot\phi=\frac{K}{f}\partial_z^2 \phi+\frac{P_g}{\omega f},
\label{e.gefdteusdb}\end{eqnarray} where $P_g$ is the power
absorption per vortex length. An important virtue of this model is
that based on our calculations of $f$ and $K$, Eq.\
(\ref{e.gefdteusdb}) predicts the time scale $L^2f/\pi^2K$ of
several minutes. Thus Eq.\ (\ref{e.gefdteusdb}) gives a simple
explanation for the observed slow mode
\cite{Dmitriev90,Sonin93,Krusius93}, which remained unexplained in
previous models \cite{Kondo91,Krusius93}.

In summary, we have investigated the spectrum of bound fermion
states localized within the vortex core of weak-coupling p-wave
superfluid. We predicted a Lifshitz transition, which
separates low-energy quasiparticle states at the half cores and
affects the rotational dynamics. Applying our results to Ref.\
\onlinecite{Kondo91} explains the observed long time scale and
thus gives one more piece of evidence of the double-core nature of
the low pressure vortex in $^3$He-B.

\begin{acknowledgments}
We thank I. Khaymovich, N. Kopnin, A. Mel'nikov, and G.
Volovik for useful discussions.
 This work was supported by the Academy of Finland and Tauno T\"onning foundation.
 M.S. and M.F. acknowledge support from the Swedish Research Council (VR).
 \end{acknowledgments}

 \clearpage
\onecolumngrid
\newcommand{\Cn}[1]{\begin{center} #1 \end{center}}
\Cn{{\Large SUPPLEMENTAL MATERIAL}\\Lifshitz transition in the double-core vortex in $^3$He-B}
\setcounter{page}{1}

 \section{Landau-Zener tunneling probability between the
quasiclassical spectrum branches}\label{e.newrotmodel}

\begin{figure}[!tb]
 \centerline{\includegraphics[width=0.7\linewidth]{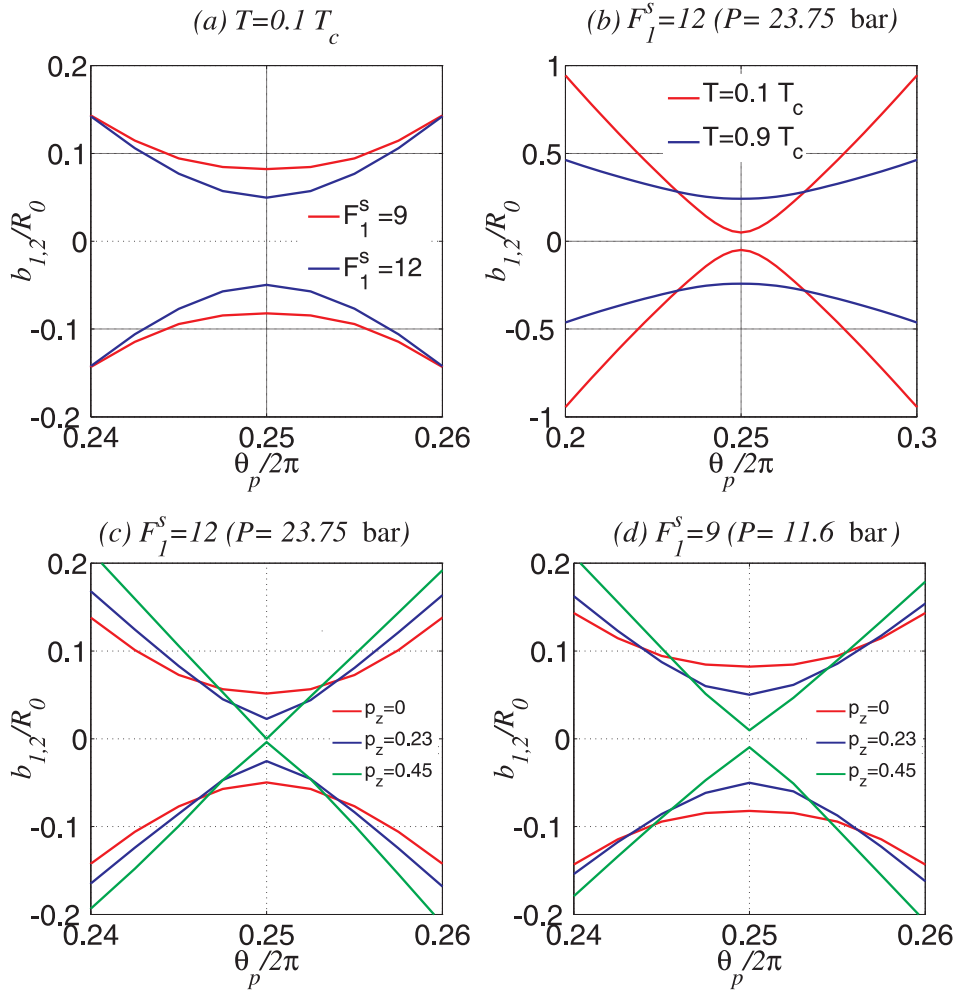}}
 \caption{\label{Fig:AntiCrossing} (Color online)
  Detailed structure of zero-energy lines $b=b_{1,2}(\theta_p)$ near the anticrossing point $\theta_p=\pi/2$
  for different values of parameters. (a) $T=0.1 T_c$ and different values
  of pressure $F_1^{\rm s}=9$ (red), $F_1^{\rm s}=12$ (blue); (b) $F_1^{\rm
  s}=12$ and $T=0.9 T_c$ (blue), $T=0.1 T_c$ (red); (c) $F_1^{\rm s}=12$, (d) $F_1^{\rm s}=9$ and
   different momentum projections $\hat p_z= 0$ (red), $0.23$ (blue), $0.45$ (green). }
 \end{figure}

 The topology of quasiparticle spectrum in double-core vortices is
 determined by the behavior of quasiclassical spectrum branches $\varepsilon= \varepsilon_{1,2}(\hat p_z,\theta_p,b)$
 near the intersection points at $\theta_p=\pi/2 + \pi n$, see
 Fig.\ 3(d) in the main text.
 In the Fig.\ \ref{Fig:AntiCrossing} we show in detail these curves in the vicinity of an anticrossing point $\theta_p=\pi/2$ for
 several values of parameters. This plot illustrates  main
 properties of the splitting $\delta b$ as function of temperature $T$, pressure $P$, momentum projection to
 the vortex axis $\hat p_z$.

 Quantitatively the splitting $\delta
 b$ is determined by the overlap between quasiparticle wave
 functions localized at different half-cores as shown in the Fig.
 1 in the main text.  At low temperatures the characteristic localization scale  is determined by the coherence length
 $\xi_0$. On the other hand the distance between vortex cores $a$ is determined by the scale $R_0= (1+F_1^{\rm s})\xi_0$ so that the ratio
 $R_0/\xi_0$ increases with growing pressure. Hence the overlap at large pressures is weaker and the splitting $\delta b $ decreases
 as can be seen comparing the curves for $F_1^{\rm s} =9$ ($P=11.6\; {\rm bar}$) and  $F_1^{\rm s} =12$ ($ P= 23.75\; {\rm bar}$) in
 Fig.\ \ref{Fig:AntiCrossing}(a). At larger temperatures the
 quasiparticle localization is determined by the temperature-dependent coherence length
 $\xi\approx \hbar v_F/\Delta$
 where $\Delta = \Delta(T)$ is the gap amplitude. On the other
 hand the distance between vortex cores for temperatures up to $T=0.99 T_c$
 has a much weaker temperature dependence for $F_1^{\rm s}=6,\;9,\;12$ as
 shown in Fig.\ 2 in the main text. Thus the overlap between
 half-core states and the splitting $\delta b$ strongly decrease with decreasing temperature
 which can be seen from the comparison of the curves
 $b=b_{1,2}(\theta_p)$ for $T=0.9 T_c$ and $0.1 T_c$ in  Fig.\ \ref{Fig:AntiCrossing}(b) .

 The most striking is the dependence of splitting $\delta b$ on
 the quasiparticle momentum projection on the vortex axis $\hat p_z$. The
  absolute value of momentum is fixed and determined by the Fermi momentum.
  Hence different values of $\hat p_z$ correspond to the
 different angles of quasiclassical trajectories with respect to the vortex
 axis. For the finite $\hat p_z$ the effective distance between half-cores {\it along the trajectories} is
elongated by the factor of $1/\sqrt{1-\hat p_z^2}$ which decreases
the overlap of localized states and suppresses the splitting
$\delta b$. As can be seen in Figs.\ \ref{Fig:AntiCrossing}(c) and \ref{Fig:AntiCrossing}(d), the
splitting can decrease to the very small values at large $\hat p_z
\sim 1$.

\begin{figure}[!tb]
 \centerline{\includegraphics[width=0.7\linewidth]{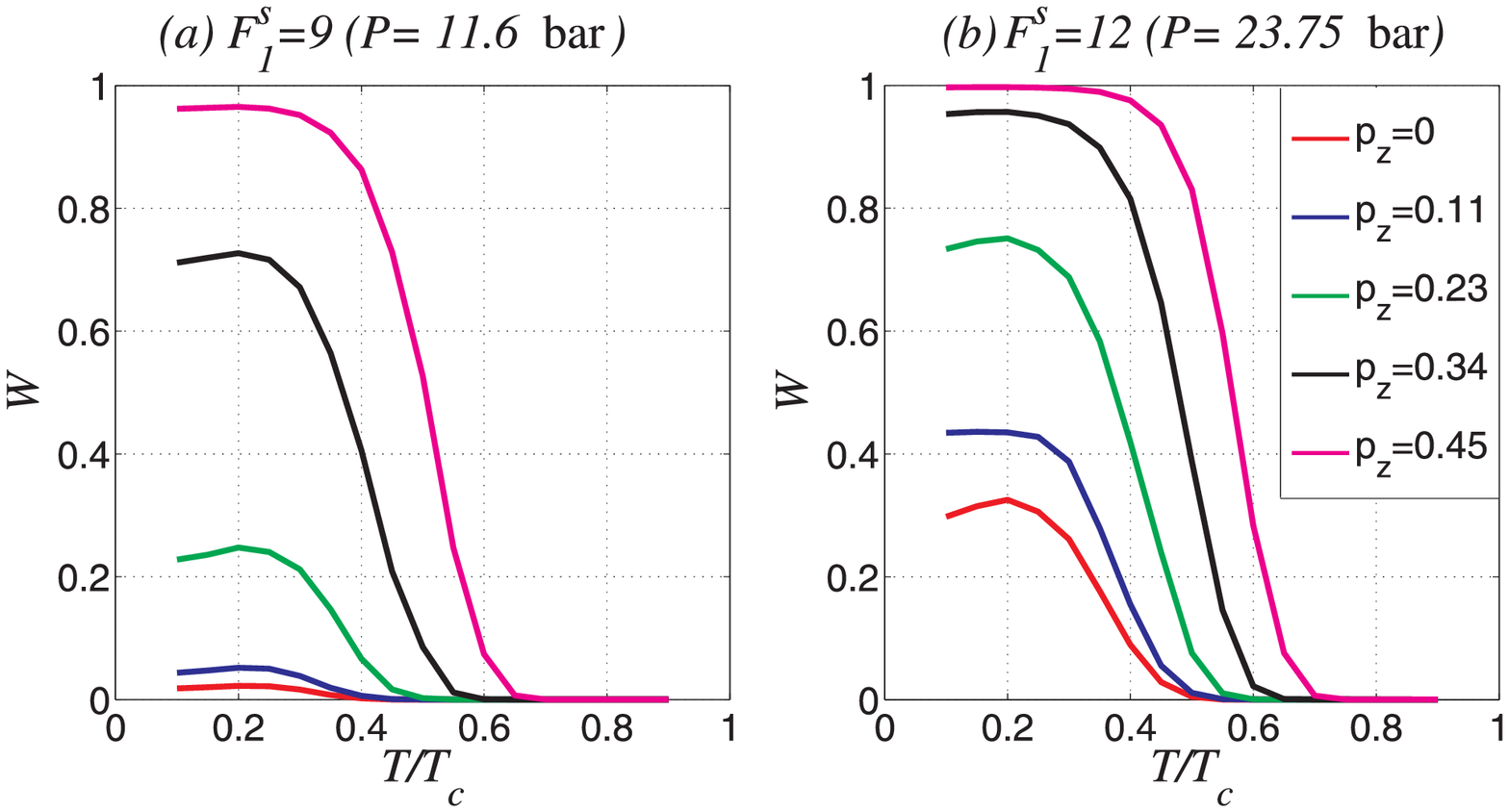}}
 \caption{\label{Fig:Probability} (Color online)
  Temperature dependencies of the Landau-Zener tunneling probability between
  quasiclassical spectral branches in the double-core vortex for (a) $F_1^{\rm s}=9$ ($P=11.6\ {\rm bar}$),
  (b) $F_1^{\rm s}=12$ ($P=23.75\ {\rm bar}$) and different values of momentum
  projection $\hat p_z$.}
 \end{figure}

 The
 splitting $\delta b$ determines the tunneling between the quasiclassical branches.
 The Landau-Zener tunneling probability $W=W (T,P,\hat p_z)$ calculated according to the general expression
 from the main text is shown in
 Figs.\ \ref{Fig:Probability}(a) and \ref{Fig:Probability}(b) as function of temperature for
 several values of $P$ and $\hat p_z$.
 These curves demonstrate a general tendency of the
 tunneling probability to increase with decreasing temperatures and increasing
 pressure. The probability is strongly enhanced at large $\hat
 p_z$. Thus  the Lifshitz transition determined by  $W\sim 1$
 occurs at different values of $P$ and $T$ for different $\hat p_z$.
 From Fig.\ \ref{Fig:Probability} one can see that at high
 pressures and low temperatures the condition $W\sim 1$ is valid
 for all values of $\hat p_z$.

\section{Model of rotational dynamics}\label{e.newrotmodel2}

The rotational dynamics of the double-core vortex was studied  by Kondo et al \cite{Kondo91}. Here we present a modified model that allows azimuthal shear of the vortex structure.  Such a case appears because  the driving dipole torque acts on the asymptotic order parameter, typically a distance $\xi_D\approx 10\ \mu$m from the vortex axis, whereas the opposing rotational friction occurs in the vortex cores, on the scale of coherence length $\xi_0\approx 10\ $nm.  Our notation follows closely Ref.\ \onlinecite{Kondo91} and the model of Ref.\ \onlinecite{Kondo91} is obtained in the limiting case $T_A\rightarrow\infty$ and $g=0$.

We study the model where the free energy of a single vortex parallel to $z$ is
\begin{eqnarray}
F=\int dz\left[-T_H(\hat S_\alpha R_{\alpha i}\hat b_i)^2+T_D(\hat{\bm a}\cdot\hat{\bm n})^2-T_A(\hat{\bm a}\cdot\hat{\bm b})^2+\frac12K(\partial_z \phi)^2\right]
\label{e.rotdyn0c}\end{eqnarray}
The magnetic field term (coefficient $T_H$) depends on the hard core anisotropy axis
$\hat{\bm b}=\hat{\bm x}\cos\phi+\hat{\bm y}\sin\phi$. The dipole term  (coefficient $T_D$) depends on the soft core anisotropy axis $\hat{\bm a}=\hat{\bm x}\cos\alpha+\hat{\bm y}\sin\alpha$. The azimuthal shear term (coefficient $T_A$) is supposed to be strong enough to keep the two anisotropy vectors nearly parallel. Other quantities are the spin-orbit rotation matrix
$R_{\alpha i}$ parametrized by axis $\hat{\bm n}=\hat{\bm y}\cos\omega t+(\hat{\bm z}\sin\eta-\hat{\bm x}\cos\eta)\sin\omega t$ and angle $\theta=\arccos(-1/4)$, the direction of  the precessing magnetization $\hat S_\alpha=R_{\alpha i}\hat H_i$, and the static field $\bm H=H(\hat{\bm x}\sin\eta+\hat{\bm z}\cos\eta)$. This model neglects the anisotropy of the dipole energy in the double-core vortex in the plane perpendicular to $\hat{\bm a}\approx \hat{\bm b}$. In principle, we should allow $\hat{\bm a}$ to have component in the $z$ direction also, but this will not affect the main results and thus is  dropped here for simplicity.
We get
\begin{eqnarray}
F=\int dz\left[-T_H(\hat {\bm H}\cdot\hat {\bm b})^2+T_D(\hat{\bm a}\cdot\hat{\bm n})^2-T_A(\hat{\bm a}\cdot\hat{\bm b})^2+\frac12K(\partial_z \phi)^2\right]\nonumber\\
=\int dz\left[-T_H\sin^2\eta\cos^2\phi+T_D(\cos\omega t\sin\alpha-\cos\eta\sin\omega t\cos\alpha)^2+T_A(\alpha-\phi)^2+\frac12K(\partial_z \phi)^2\right]
\label{e.rotdyn0b}\end{eqnarray}
Note that for $\hat{\bm H}=\hat{\bm x}$, $\cos\eta=0$ and both field and dipole terms are minimized by $\phi=\alpha=n\pi$, that is $\hat{\bm a}=\hat{\bm b}=\pm\hat{\bm x}$ and $\hat{\bm n}$ rotating around it.

We suppose the frictional equations of motion
\begin{eqnarray}
g\dot\alpha=-\frac{\delta F}{\delta \alpha},\quad f\dot\phi=-\frac{\delta F}{\delta \phi}.
\end{eqnarray}
We obtain
\begin{eqnarray}
g\dot\alpha=-\frac{\delta F}{\delta \alpha}
=-T_D\sin^2\omega t\sin^2\eta\sin2\alpha-T_D\cos2\omega t(1-\cos\eta)\sin2\alpha
+T_D\cos\eta\sin2(\omega t-\alpha)-2T_A(\alpha-\phi).
\label{e.gefdtec}\end{eqnarray}
Using the same approximation as Kondo et al this simplifies to
\begin{eqnarray}
g\dot\alpha=-\frac12T_D\sin^2\eta\sin2\alpha
+T_D\cos\eta\sin2(\omega t-\alpha)-2T_A(\alpha-\phi).
\label{e.Kondomau}\end{eqnarray}
This expression is exact in second (but not fourth) power in $\eta$. The other equation is
\begin{eqnarray}
f\dot\phi=-\frac{\delta F}{\delta \phi}=-T_H\sin^2\eta\sin2\phi+2T_A(\alpha-\phi)+K\partial_z^2 \phi.
\label{e.gefdteu}\end{eqnarray}

Suppose now that $f$ is very large, which means $\phi(t)$ is so slow that it can be taken as a constant during one cycle.
We study
\begin{eqnarray}
g\dot\alpha=-M\sin2\alpha+G\sin2(\omega t-\alpha)-2T_A(\alpha-\phi)
\end{eqnarray}
where in the approximation above $M=\frac12T_D\sin^2\eta$ and $G=T_D\cos\eta$.
Suppose $\alpha(t)=\alpha_0+\alpha_1(t)$ with small $\alpha_1$
\begin{eqnarray}
g\dot\alpha_1
=-M\sin2\alpha_0-2T_A(\alpha_0-\phi)+G\sin2(\omega t-\alpha_0)-2(T_A+M\cos2\alpha_0)\alpha_1
-2G\cos 2(\omega t-\alpha_0)\alpha_1.
\end{eqnarray}
Assuming $\alpha_1=A\cos 2(\omega t-\chi)$
\begin{eqnarray}
-2\omega g A\sin 2(\omega t-\chi)\nonumber\\=-M\sin2\alpha_0-2T_A(\alpha_0-\phi)+G\sin2(\omega t-\alpha_0)-2A(T_A+M\cos2\alpha_0)\cos 2(\omega t-\chi)\nonumber\\
-GA\cos 2(2\omega t-\alpha_0-\chi)-GA\cos 2(\chi-\alpha_0).
\end{eqnarray}
At frequencies $0$ and $2\omega$ we get
\begin{eqnarray}
0=-M\sin2\alpha_0-2T_A(\alpha_0-\phi)-GA\cos 2(\chi-\alpha_0)\nonumber\\
-2\omega g A\sin 2(\omega t-\chi)=G\sin2(\omega t-\alpha_0)-2A(T_A+M\cos2\alpha_0)\cos 2(\omega t-\chi)
\label{e.rodvep}\end{eqnarray}
The latter equation can be written
\begin{eqnarray}
2A(T_A+M\cos2\alpha_0)\cos 2(\omega t-\chi)-2\omega g A\sin 2(\omega t-\chi)=G\sin2(\omega t-\alpha_0)
\end{eqnarray}
\begin{eqnarray}
2A(T_A+M\cos2\alpha_0)\cos 2\omega t-2\omega g A\sin 2\omega t=G\sin2(\omega t+\chi-\alpha_0)\nonumber\\
=G[\sin2\omega t\cos2(\chi-\alpha_0)+\cos2\omega t\sin2(\chi-\alpha_0)]
\end{eqnarray}
\begin{eqnarray}
2A(T_A+M\cos2\alpha_0)=G\sin2(\chi-\alpha_0)\nonumber\\
-2\omega g A=G\cos2(\chi-\alpha_0)
\label{e.2ATSmcos}\end{eqnarray}
\begin{eqnarray}
A^2=\frac{G^2/4}{\omega^2 g^2+(T_A+M\cos2\alpha_0)^2}\approx \frac{G^2/4}{\omega^2 g^2+T_A^2}
=\frac{\frac14T_D^2\cos^2\eta}{\omega^2 g^2+T_A^2}.
\label{e.A^2=fracG}\end{eqnarray}
The absorbed power is determined by $A$,
\begin{eqnarray}
P_g=-\langle\dot\alpha\frac{\delta F}{\delta \alpha}\rangle=g\langle\dot\alpha^2\rangle=2g\omega^2A^2.
\label{e.P_g=-N_v}\end{eqnarray}
Note that here $P$ denotes the absorption per vortex length whereas in Ref.\ \onlinecite{Kondo91} $P$ denotes the total absorption.

Applying the latter of (\ref{e.2ATSmcos}) to the first equation (\ref{e.rodvep}) gives
\begin{eqnarray}
M\sin2\alpha_0+2T_A(\alpha_0-\phi)-2\omega gA^2=0.
\label{e.rodvex}\end{eqnarray}
From (\ref{e.gefdteu}) we have
\begin{eqnarray}
f\dot\phi=-T_H\sin^2\eta\sin2\phi+2T_A(\alpha_0-\phi)+K\partial_z^2
\phi. \label{e.gefdteusd}\end{eqnarray} Because $\phi$ is a slow
variable, it is only the low frequency limit of $f$ that appears.
What remains to be solved is $\phi$ and $\alpha_0$ from equations
(\ref{e.rodvex}) and (\ref{e.gefdteusd}). Two different cases need
to be studied.

1) Rocking oscillations. We assume a solution where the left hand side of  (\ref{e.gefdteusd}) vanishes,
\begin{eqnarray}
-\frac{\delta F}{\delta \phi}=-T_H\sin^2\eta\sin2\phi+2T_A(\alpha_0-\phi)+K\partial_z^2 \phi=0.
\end{eqnarray}
Supposing $\phi-\alpha\ll 1$ we can solve this
\begin{eqnarray}
\phi=\alpha_0-\frac{T_H\sin^2\eta\sin2\alpha_0-K\partial_z^2 \alpha_0}{2(T_H\sin^2\eta\cos2\alpha_0+T_A)}\approx\alpha_0-\frac{T_H\sin^2\eta\sin2\alpha_0-K\partial_z^2 \alpha_0}{2T_A}.
\label{e.phi=alpha_0}\end{eqnarray}
The lag of $\phi$ behind $\alpha_0$ is an increasing function of $\phi\approx\alpha_0$ for $0<\phi\approx\alpha_0<\pi/4$.
Substituting (\ref{e.phi=alpha_0}) into (\ref{e.rodvex}) gives
\begin{eqnarray}
(T_H+\frac12T_D)\sin^2\eta\sin2\alpha_0-K\partial_z^2 \alpha_0-2\omega gA^2=0
\label{e.rodve3}\end{eqnarray}
and thus
\begin{eqnarray}
\sin2\alpha_0=\frac{2\omega gA^2+K\partial_z^2 \alpha_0}{(T_H+\frac12T_D)\sin^2\eta}.
\label{e.rodve4}\end{eqnarray}
This solution is possible only if the right hand side of (\ref{e.rodve4}) is less than unity. At equality, $\alpha_0=\pi/4+n\pi$. The stability condition at $\partial_z^2 \alpha_0=0$ can be written
\begin{eqnarray}
\tan^2\eta>\tan^2\eta_0=\frac{\frac12\omega g T_D^2}{(T_H+\frac12T_D)(\omega^2 g^2+T_A^2)}.
\end{eqnarray}
Interestingly, this can also be written
\begin{eqnarray}
\sin^2\eta>\sin^2\eta_0=\frac{P_g}{\omega (T_H+\frac12T_D)}.
\end{eqnarray}
These results are essentially ($\tan \eta_0\approx \sin\eta_0$) the same as in Ref.\ \onlinecite{Kondo91} if one makes the replacement
$f\rightarrow g+T_A^2/g\omega^2$.

2) In the case solution (\ref{e.rodve4}) is not possible, we get slow rolling motion of $\phi(t)$. For simplicity we consider $\eta=0$ only.
From (\ref{e.rodvex}) we get
\begin{eqnarray}
2T_A(\alpha_0-\phi)=2\omega gA^2
\label{e.rodvexb}\end{eqnarray}
and substitution to (\ref{e.gefdteusd}) gives
$$
\dot\phi=\frac{K }{f}\partial_z^2\phi+\frac{P_g}{\omega f}.\eqno{(\ref{e.gefdteusdb})}
$$

Equation (\ref{e.gefdteusdb}) is a diffusion equation
\begin{eqnarray}
\dot\phi=D\partial_z^2\phi +C
\label{e.gefdteuswie}\end{eqnarray}
The solution for $0<z<L$ and $t>0$ with boundary conditions $\phi(z=0)=0=\phi(z=L)$ can be found as Fourier series
\begin{eqnarray}
\phi=\frac{C}{2D}(Lz-z^2)+\sum_n B_ne^{-Dk_n^2t}\sin k_nz
\end{eqnarray}
with $k_n=n\pi/L$. The slowest component has the rate
\begin{eqnarray}
\Gamma=\pi^2 D/L^2=\pi^2 K/fL^2.
\label{e.diferrat}\end{eqnarray}

We compare our results to Ref.\ \onlinecite{Kondo91}. For that we
give numerical values of our  weak-coupling results corresponding
to 29.3 bars pressure, $T/T_c=0.5$ and $L=7$ mm. At this pressure
$F_1^s=13.3$ and we neglect other Fermi-liquid interactions. For
the friction coefficient we get $f\approx2\times10^{-19}$ J s/m,
see Section \ref{s.rotfr}. In section \ref{s.twist} we calculate
$K\approx5\times10^{-27}$ Jm. This gives the time constant
$\Gamma^{-1}\approx3$ min. This order of magnitude is in good
agreement with the measured value of the ``slow mode'' in Ref.\
\onlinecite{Dmitriev90,Sonin93,Krusius93}. Extensive measurements of the slow mode
are presented in Ref.\ \onlinecite{Krusius93}. It seems that these
results need to be reinterpreted using the diffusion equation
(\ref{e.gefdteusdb}).

Our results could be compared to $f=4.5\times10^{-23}$ J s/m and $K=1.1\times10^{-25}$ Jm that were obtained by fitting the experiment to the model in Ref.\ \onlinecite{Kondo91}.

\section{Rotational friction}\label{s.rotfr}

The calculation of the rotational friction coefficient $f$ in the
main text applies the kinetic equation of Ref.\
\onlinecite{RotatingVortex} to our numerical solution of the
excitation spectrum in the vortex core. The results are presented
in Fig.\ \ref{Fig:Friction}. Evaluating this at the conditions of
Ref.\ \onlinecite{Kondo91} (see above) gives $f=1.7\times10^{-19}$
J s/m in the low frequency limit. The result at the Larmor
precession frequency at $H=14.2$ mT ($\hbar\omega/E_m=3.9$) would
be $f(\omega_L)= 0.9\times10^{-19}$ J s/m.

An alternative estimate of $f$ is obtained as follows.  In the
case of strong Landau-Zener tunneling, the half cores behave as
separated half-quantum vortices except that they are bound
together at the distance $a$. The standard mutual friction force
\cite{BevanJLTP1997,KopninReview} applied to both half cores gives
\begin{eqnarray}
f=\frac14a^2\kappa\rho_sd_\parallel,
\label{e.ffshqvp}\end{eqnarray} where $\kappa=\pi\hbar/m$ is the
circulation quantum, $\rho_s$ the superfluid density and
$d_\parallel$ the mutual friction parameter. Taking $a$ from Fig.\
\ref{Fig:OrderParameterFermi} and $d_\parallel$ from Ref.\
\onlinecite{BevanJLTP1997} gives $f=1.6\times10^{-19}$ J s/m at
the conditions stated above.

\section{Twist rigidity}\label{s.twist}

The twisting of a  double-core vortex is measured by a twist wave vector $k=\partial_z\phi$. For a small twist the additional free energy is supposed to be quadratic in $k$,
\begin{eqnarray}
F_{\rm twist}={\textstyle\frac12}Kk^2.
\label{e.twscofde}\end{eqnarray}
Here we aim to calculate the twist-rigidity coefficient $K$.

The  reduced order parameter is expected to satisfy the boundary condition $\tilde{\textsf{A}}(r,\varphi,z)\rightarrow \textsf{I}e^{i\varphi}$ for $r\rightarrow \infty$, where $\textsf{I}$ is a unit matrix. An axially symmetric vortex satisfies the following symmetry \cite{Thuneberg-PRB}
\begin{eqnarray}
\tilde{\textsf{A}}(r,\phi)=e^{i\theta}\textsf{R}(\hat{\bm z},\theta)\tilde{\textsf{A}}(r,\phi-\theta)\textsf{R}(\hat{\bm z},-\theta),
\end{eqnarray}
where $\textsf{R}$ is a rotation matrix parametrized by an axis and angle of rotation. A twisted vortex apparently has the property
\begin{eqnarray}
\tilde{\textsf{A}}(r,\phi,z)=e^{ikz}\textsf{R}(\hat{\bm z},kz)\tilde{\textsf{A}}(r,\phi-kz,0)\textsf{R}(\hat{\bm z},-kz).
\label{e.vtwiop}\end{eqnarray}
Taking the $z$ derivative and evaluating it at $z=0$ gives
\begin{eqnarray}
\frac{\partial\tilde{\textsf{A}}}{\partial z}=k\textsf{B},
\end{eqnarray}
where
\begin{eqnarray}
\textsf{B}=i\tilde{\textsf{A}}+\textsf{R}'\tilde{\textsf{A}}-\tilde{\textsf{A}}\textsf{R}'+y\frac{\partial\tilde{\textsf{A}}}{\partial x}-x\frac{\partial\tilde{\textsf{A}}}{\partial y},\quad \textsf{R}'=\left(\begin{array}{ccc}0 & -1 & 0 \\1 & 0 & 0 \\0 & 0 & 0\end{array}\right).
\end{eqnarray}
In the Ginzburg-Landau region the twist-rigidity coefficient $K$ is given by
\begin{eqnarray}
K=2\lambda_{\rm G2}\int d^2r\sum_{\mu=1}^3\left[\sum_{j=1}^3|B_{\mu j}|^2+2|B_{\mu z}|^2\right].
\label{e.twckge}\end{eqnarray}
In order to get a good approximation of the twist rigidity at general temperature, we have written the prefactor using \cite{Thuneberg01}
\begin{equation}
\lambda_{\rm G2}=\frac{\hbar^2\rho(1-Y)}{ 40mm^*} .
\label{e.rhosfkw}\end{equation}
This form of  $\lambda_{\rm G2}$ is valid since we assume  $F_1^{\rm a}=F_3^{\rm a}=0$.
Physically, the  twisting (\ref{e.vtwiop}) causes axial spin flow in the asymptotic region and axial mass flow in the vortex center, but the $F_1^{\rm a}$ dependence of the former and the $F_1^{\rm s}$ dependence of the  latter are not properly included in (\ref{e.twckge}).

The asymptotic form of the order parameter (neglecting dipole-dipole coupling) far from the vortex axis is of the form $\tilde{\textsf{A}}={\textsf{R}}(\bm\theta)+O(r^{-2})$, where
\begin{eqnarray}
\bm\theta=\frac{C_1\cos\phi}r\left(\frac{\sin\phi}{1+c}\hat{\bm r}+\cos\phi\hat{\bm \phi}\right)
+\frac{C_2\sin\phi}r\left(-\frac{\cos\phi}{1+c}\hat{\bm r}+\sin\phi\hat{\bm \phi}\right).
\label{e.vtaildeso2}\end{eqnarray}
In general  $c=\lambda_{\rm G1}/2\lambda_{\rm G2}$ but in the present approximation $F_1^{\rm a}=F_3^{\rm a}=0$ we have $c=1$ \cite{Thuneberg01}.
Based on this, we can calculate the asymptotic contribution to the twist rigidity.  Since (\ref{e.vtaildeso2})  is expressed in cylindrical coordinates, the twist energy can be calculated more directly than above, by replacing
$\phi$  by $\phi-kz$. The additional energy can be evaluated from the gradient energy \cite{Thuneberg01}
\begin{eqnarray}
F_{Gz}=2\lambda_{\rm G2}(1+c)\int d^2r\frac{\partial \theta_k}{\partial z}\frac{\partial \theta_k}{\partial z}
=2\lambda_{\rm G2}(1+c)k^2\int d^2r\frac{\partial \theta_k}{\partial \phi}\frac{\partial \theta_k}{\partial \phi}\nonumber\\
=2\lambda_{\rm G2}(1+c)k^2\int d\phi\int dr\frac1r\left[\frac{(\cos^2\phi-\sin^2\phi)^2}{(1+c)^2}+4\cos^2\phi\sin^2\phi\right](C_1-C_2)^2\nonumber\\
=2\pi\lambda_{\rm G2}k^2\frac{2+2c+c^2}{1+c}(C_1-C_2)^2\int dr\frac1r.
\label{e.fhygrt2}\end{eqnarray}
Comparison to (\ref{e.twscofde}) allows to identify
\begin{eqnarray}
K_{\rm tail}=4\pi\lambda_{\rm G2}\frac{2+2c+c^2}{1+c}(C_1-C_2)^2\int dr\frac1r.
\label{e.fhygrt3a}\end{eqnarray}
The dipole-dipole energy suppresses the contribution at distances beyond the dipole length $\xi_d$.

We estimate numerical values under conditions explained above
[below Eq.\ (\ref{e.diferrat})]. Substituting the numerical order
parameter into (\ref{e.twckge}) gives $K=3.2\times10^{-27}$ Jm  in
the region $r<60\xi_0$. Adding the asymptotic part from
(\ref{e.fhygrt3a}) with $C_1= 3.7R_0$ and $C_2=0.1R_0$ we get
$K=4.0\times10^{-27}$ Jm.

Another estimation of the twist rigidity is to calculate the twisting energy of a pair of half-quantum vortices. Estimating energy by the length increase of the twisted pair gives
\begin{eqnarray}
K_{\rm mass\ flow}=\frac{\rho_s\kappa^2 a^2}{32\pi}\ln\frac{a}{r_c}
\end{eqnarray}
where $a$ is the distance between the half cores and $r_c$ their radius. Estimating $a=5.4R_0$ gives
$K_{\rm mass\ flow}=1.9\times10^{-27}$ Jm. Estimating this without Fermi-liquid interaction, i.e. as we did in (\ref{e.twckge}), we would get $K_{\rm mass\ flow0}=0.6\times10^{-27}$ Jm. Supposing this difference is just missing in the formula based on (\ref{e.twckge}), the corrected value for total twist coefficient is
$K=5.3\times10^{-27}$ Jm.



 \end{document}